# Short-range Magnetic Ordering in the Geometrically Frustrated Layered Compound YBaCo$_4$O$_7$ with an Extended Kagomé Structure

**A. K. Bera [a], S. M. Yusuf [a*], and S. Banerjee [b]**

[a]*Solid State Physics Division, Bhabha Atomic Research Centre, Mumbai 400 085, India*
[b]*Department of Atomic Energy, Mumbai 400 001, India*

**Abstract:** Structural and magnetic properties of the geometrically frustrated layered compound YBaCo$_4$O$_7$ have been studied by magnetization and neutron diffraction. A layered type crystal structure, with alternating tetrahedral layers of Kagomé and triangular types, along the *c* axis (hexagonal symmetry, space group *P*6$_3$*mc*) has been found. The oxygen content of the present compound was determined by iodometric titration to be 7.00±0.01. Presence of a short-range antiferromagnetic ordering (below $T_N$ ~ 110 K) has been concluded from the appearance of an addition broad magnetic peak (at $Q$ ~ 1.35 Å$^{-1}$) in the neutron diffraction pattern. A staggered chiral type ($\sqrt{3} \times \sqrt{3}$) spin structure in the Kagomé layers has been found. Presence of a significant magneto-structural coupling is concluded from the changes in lattice parameters across the magnetic ordering temperature $T_N$ ~ 110 K.

*Corresponding author:

Email address: smyusuf@barc.gov.in (S. M. Yusuf), FAX: +91 22 25505151, Telephone: +91 22 25595608

## 1. Introduction

Layered cobaltites with chemical formula $R$BaCo$_4$O$_{7+\delta}$ ($R$ = Y, and rare-earth ions), a new class of geometrically frustrated magnets, have attracted a lot of attention in the recent years due to their unusual physical properties [1-12]. The main interesting feature of these compounds is their peculiar crystal structure which leads to the unusual physical properties. The crystal structure of the $R$BaCo$_4$O$_7$ compounds can be described as a stacking of CoO$_4$ tetrahedral layers along the crystallographic $c$-direction. The alternative tetrahedral layers are Kagomé and triangular types. A spin-frustration with different types of magnetic correlations within the individual layers is expected in these compounds due to the Kagomé and triangular types of geometrical arrangements of the magnetic ions in the alternating layers. The magnetic correlations are also expected to remain confined within the Co layers in the $ab$-plane (*i.e*, two dimensional (2D) magnetic correlation) due to the peculiar layered type crystal structure of this system. However, the inter-planar coupling in the third direction ($c$-axis), if any, might play an important role to set in a three dimensional magnetic ordering in such compounds. Different types of magnetic orderings such as short-range antiferromagnetic (AFM) ordering [1, 2, 5, 12], 3D long-range AFM ordering [4, 13], field induced weak ferromagnetic ordering [9], *etc.* have been reported for different compounds of the series $R$BaCo$_4$O$_{7+\delta}$ with different $R$ ions and oxygen content. These compounds also show a large thermoelectric power factors at higher temperature and can be suitable for the thermoelectric power generation [8]. Moreover, these compounds show a high degree of chemical flexibility, having a large capacity for reversible oxygen absorption and desorption, which makes these materials suitable for applications in oxygen storage, oxygen sensors, oxygen membranes, solid-oxide fuel cells, *etc* [14]. Understanding of the magnetic ground state of this special class of compounds and its relation with the complex crystal structure is an important and fundamental issue in recent years.

In the present study, we are particularly interested in the Y based layered cobaltites e.g., $YBaCo_4O_7$. A number of efforts have been made in the recent years by several researchers to understand the complex magnetic properties of the compound $YBaCo_4O_{7+\delta}$ and there exists a lot of controversies on magnetic ordering as well as crystal structure for this compound [1-5, 15-18]. A room temperature hexagonal crystal structure with $P6_3mc$ space group was reported by several groups [1-3, 5, 15-17]. On the other hand, a trigonal crystal structure (space group $P31c$) with a first order phase transition to an orthorhombic crystal structure (space group $Pbn2_1$) at lower temperature ($T \leq 313$ K) was also reported [4, 18]. Form magnetic study, the presence of a broad magnetic peak in the powder neutron diffraction pattern, corresponding to a short-range spin-spin correlation length of 4.2 - 4.6 Å, at 10 K, was reported by Valldor $et\ al.$ [1]. The magnetic ground state was defined as a disordered antiferromagnet. In agreement with the powder neutron diffraction data, the single crystal neutron diffraction study by Soda $et.\ al.$ [5] also showed an absence of a long-range magnetic ordering down to 10 K. On the contrary, a magnetic Bragg scattering due to a long-range magnetic ordering below 110 K was reported by Chapon $et\ al.$[4]. In addition to the magnetic Bragg scattering, a diffuse magnetic scattering over a wide temperature range above 110 K was also reported and interpreted as arising from an independent short-range magnetic ordering of both Kagomé and triangular layers [4]. Recent Mössbauer study by Tsipis $et\ al.$ [7] on 1% Fe doped compound $YBaCo_{3.96}Fe_{0.04}O_{7.02}$ revealed a gradual freezing of the iron magnetic moments with decreasing temperature below ~ 75–80 K. Besides these controversial claims, there is no report, so far, on the dimensionality of short-range magnetic correlation as well as temperature variation of the spin-spin correlation length. It is, therefore, quite interesting and important to carry out a detailed magnetic study of the $YBaCo_4O_7$ compound to address the nature of the magnetic ground state. In addition, generally a magneto-structural coupling is expected in such types of geometrically frustrated systems. Therefore, a detailed microscopic study on the temperature dependence of the crystal structure and magnetic correlation is essential for this system. In this paper, we report the results of detailed structural as well as magnetic properties study on a stoichiometric compound of $YBaCo_4O_7$ by employing neutron diffraction and dc magnetization techniques. Here, we have carried out neutron diffraction study as a function of temperature over a broad range to bring out a temperature evolution of the crystal structure as well as magnetic ordering, and the coupled structural and magnetic correlations. An attempt has been made to understand the dimensionality of the magnetic ordering in this compound. The temperature dependence of the spin-spin correlation length has been studied as well. Our study reveals that the system crystallizes in the hexagonal symmetry with $P6_3mc$ space group. An antiferromagnetic (AFM) ordering has been observed below 110 K. It has also been observed that the spin-spin correlation length remains short-range down to 22 K, the lowest measured temperature. The role of crystal structure on the magnetic ordering has been brought out.

## 2. Experimental

Polycrystalline sample of $YBaCo_4O_7$ was synthesized by the conventional solid state reaction method. Stoichiometric amounts of high purity ($\geq 99.99\%$) $Y_2O_3$, $BaCO_3$, and $Co_3O_4$ were initially mixed using an agate mortar pestle and placed in a platinum crucible. The well ground powders were decarbonated at 1000 °C for 20 hours with intermediate grindings and then pressed into a pellet form. The palletized mixtures were heated at 1200 °C in air for total 60 hours with intermediate grindings.

The phase purity of the sample was ensured by the room temperature powder x-ray diffraction study using a Cu $K_\alpha$ radiation over the scattering angular range $10° \leq 2\theta \leq 90°$ (not shown). The sample was found to be single phase in nature. The oxygen stoichiometry of the compound was examined by an iodometric titration, and experimentally verified to be nearly stoichiometric (7.00±0.01) within the experimental limits.

DC magnetization measurements were carried out using a commercial vibrating sample magnetometer (VSM). The zero field cooled magnetization ($M_{ZFC}$) and field cooled magnetization ($M_{FC}$) were measured under 1 kOe field over the temperature range 5-310 K. The $M_{ZFC}$ and $M_{FC}$ measurements were carried out in warming cycles after cooling the sample in the absence of field and in the presence of field (1 kOe), respectively. The isothermal magnetization was measured at different temperatures down to 5 K (cooled under zero field cooled condition) in the magnetic field range of ± 70 kOe.

Neutron powder diffraction measurements were carried out by using the five linear position sensitive detector (PSD) based powder diffractometer II ($\lambda$ =1.249 Å) at Dhruva research reactor, Trombay, India at selected temperatures between 22 and 300 K over 2θ range of 5° to 138° covering a $Q$ (= 4πsinθ/$\lambda$) range of 0.35 to 9.4 Å$^{-1}$. The powdered sample was packed in a cylindrical vanadium container and attached to the cold finger of a closed cycle helium refrigerator. The diffraction data were analyzed by the Rietveld method using the FULLPROF program [19].

## 3. Results and discussion

### 3.1. Magnetization Study

The temperature dependent $M_{ZFC}$ and $M_{FC}$ curves are shown in Fig. 1. With lowering of temperature, $M_{ZFC}$ ($T$) curve shows a sharp peak at a temperature ($T_P$) ~ 70.7 K and followed by a broad hump over the temperature range of 60-15 K. On the other hand, the $M_{FC}$ curve shows a sharp rise at ~ 80 K with a broad peak over 80-20 K centering at ~ 50 K. A bifurcation in the $M_{ZFC}$ and $M_{FC}$ curves has been observed below ~ 80 K. A similar type of temperature dependent behavior for the present compound (in its polycrystalline form) was reported by Valldor *et al.* [1] as well as for the single crystal by Soda *et al.* [5]  For the single crystal, the anomaly was reported only for the planner (*ab* plane) susceptibility and was argued that this was due to the presence of the Co moments within the *ab* plane [5]. Although, the magnetization curves show a strong anomaly only below ~ 80 K, our neutron study shows that an onset of magnetic ordering occurs at a much higher temperature ~ 110 K (discussed later).  At ~ 110 K, a weak slope change has been observed in both $M_{ZFC}(T)$ and $M_{FC}(T)$ curves [Fig. 1 and its inset], which can be taken as a signature of the onset of the magnetic ordering.  The observed anomaly in the $M_{ZFC}(T)$ and $M_{FC}(T)$ curves at ~ 70 K may be due to a reorientation  of the magnetic moments associated with a structural transition as reported from a single crystal study by Soda *et al* [5]. However, in the present powder neutron diffraction study, we have not found any such structural transition at ~ 70 K (discuss later). In the paramagnetic state (110 K $\leq T \leq$ 300 K), the susceptibility curve (shown in inset of Fig. 1) does not follow the Curie-Weiss law. However, an indication of negative Curie-Weiss temperature ($\theta_{C-W}$) is evident which suggests the presence of a very short-range AFM correlation within both Kagomé and triangular layers over a broad temperature range. The presence of strong antiferromagnetic interactions, with a large Curie-Weiss temperature ($\theta_{C-W}$ ~ −907 K) was concluded in literature [4, 15]. The very high value of the $\theta_{C-W}$ as compared to the actual magnetic ordering temperature ($T_N$ ~ 110 K) is a signature of a geometrical spin frustration [20-22].

Figure 2 shows the isotherm magnetization curves as a function of magnetic field at a few selective temperatures (5, 40, 90, and 130 K), below and above the magnetic ordering temperature (~ 110 K). A linear magnetic field dependent magnetization behavior has been observed at all temperatures, suggesting the presence of an AFM type correlation for this compound. The inset of Fig. 2 shows the $M(H)$ curves over all four quadrant at 5 K.  No hysteresis has been found at this temperature. It may be noted that a spin-glass magnetic state with a weak ferromagnetic component below 40 K was reported for a hyper-stoichiometric sample $YBaCo_4O_{8.5}$ where a clear hysteresis in the $M(H)$ curve was found at 4.2 K [6]. This indicates that the magnetic behavior of $YBaCo_4O_7$ strongly depends on the oxygen stoichiometry of the sample because the ratio between the oxidation states of the Co ions, *i.e.*, $Co^{2+}:Co^{3+}$ is directly related to the oxygen content in this compound. The compound, studied by us, is fully stoichiometric (7.00±0.01) as confirmed by the iodometric titration analysis.

### 3.2   *Neutron Diffraction Study*

Figure 3 shows the Rietveld refined neutron diffraction patterns at 300, 70, and 40 K for $YBaCo_4O_7$. The refinement confirms that the compound crystallizes in the hexagonal symmetry with the space group $P6_3mc$ at all temperatures as reported earlier [1-3, 5]. There is some debate on the correct space group, between $P6_3mc$ (No. 186) and $P31c$ (No. 159), of the high temperature phase of $YBaCo_4O_7$ and other compounds of the $RBaCo_4O_7$ series. Recently, Avdeev *et al.* [23] studied the stability of the crystal structure of the $RBaCo_4O_7$ cobaltites based on the bond-valence calculation. In the context of their analysis, it was reported that both space groups lead to the same result and therefore, the higher symmetry group $P6_3mc$ can be considered. In our case, fits with almost equal quality were obtained with both space groups ($\chi^2$ = 2.34% for $P6_3mc$ and $\chi^2$ = 2.35% for $P31c$) at 300 K. Therefore, we have considered the $P6_3mc$ space group for all subsequent analysis. In addition to the neutron diffraction data, an electron diffraction study by Valldor *et al*. [1] also confirmed the space group $P6_3mc$. The absence of (0 0 *l*) reflections in the electron diffraction pattern with odd values of *l* confirmed the presence of $6_3$ screw axis. On the other hand, an oxygen stoichiomety in the present compound may stabilize the higher symmetry ($P6_3mc$) crystal structure, as reported recently for the isostructural compound $YBaCo_3AlO_7$ [3]. For the studied compound, the oxygen stroichiometry (7.00±0.01) was confirmed by us from the iodiometric titration study. The derived values of unit cell parameters, unit cell volume, fractional atomic coordinates, and isotropic temperature factors are given in Table 1. The Rietveld refinement suggests a full occupancy (100%) of the ions at all seven crystallographic sites and, therefore, the site occupancies are kept fixed to 1 during the refinement. The layered type crystal structure for the present compound $YBaCo_4O_7$ is shown in Fig. 4(a). The geometrical arrangements of two alternative layers *i.e.*, triangular (2*a* site) and Kagomé (6*c* site) layers are shown in Figs. 4 (b) and 4 (c), respectively. It is important to mention that the geometry of the Kagomé lattice at the 6c site is not a perfect Kagomé type. Rather, this is a distorted Kagomé lattice formed by two different sizes of corner sharing equilateral triangles (one is connected via O1 and other one is connected via O2) [Fig. 4 (c)]. A perfect Kagomé lattice is possible in this hexagonal crystal structure for a general position of type (*x*, 1-*x*, z) only with *x* = 1/6. However, the value of *x* for the present compound is found to be 0.1764(9) at 300 K. The side lengths of these equilateral triangles in the Kagomé plane are found to be 2.953(2) Å and 3.319(2) Å, respectively at 300 K. On the other hand, triangular layers are formed by side sharing equilateral triangles. Here, the Co ions are well separated, and the distance (Co-Co separation) is found to be 5.10(3) Å at 300 K. The alternative tetrahedral layers are occupied by 75% (Kagomé layer) and 25% (triangular layer) of the cobalt cations. The Y ions are situated within the triangular layers, whereas, the Ba ions occupy the positions in between Kagomé and triangular layers. Three different types of oxygen coordinations viz. an anticuboctahedron [Coordination Number (CN) = 12] for Ba ions (2*b* site), an octahedron (CN = 6) for Y ions (2*b* site), and a tetrahedron (CN = 4) for Co ions (2*a* and 6*c* sites) are found. Three different oxygen sites are labeled as O1, O2, and O3. In a given Kagomé layer, the $CoO_4$ tetrahedra are connected by sharing their corners via O1 and O2 oxygen ions as shown in Fig. 4 (c). On the other hand, in a given triangular layer [Fig. 4 (b)], the $CoO_4$ tetrahedra are well separated and connected via $YO_6$ or/and $BaO_{12}$ polyhedrons. As a result, the superexchange pathways between Co ions, through Co-O-Y/Ba-O-Co network, are quit longer. Therefore, a weak in-plane magnetic interaction between Co ions is expected within the triangular layer as compared to that in the Kagomé layer, where the Co ions are connected via oxygen ions (O1 and O2) alone. Along the perpendicular direction, the Kagomé and triangular layers are connected by sharing corners via O3 and O2 oxygen ions [Fig. 4 (d)]. The refined values of bond lengths between metal ions and oxygen ions are given in Table 2. The present system shows a complex crystal structure where a magneto-structural coupling can be expected (discussed later).

Figure 5 (a) shows the observed neutron diffraction pattern recorded at 22 K, and a calculated pattern by considering only the nuclear (crystal structure) phase. An appearance of an additional broad (not instrumental resolution limited) asymmetric type peak at $Q \sim 1.35$ Å$^{-1}$ suggests the presence of an AFM correlation at this temperature. This also suggests that a long-range magnetic ordering is absent in the present compound. It has been observed that the peak intensity (at $Q \sim 1.35$ Å$^{-1}$) increases gradually below $\sim 110$ K without any saturation down to 22 K [Fig. 5 (b) and its insets]. Interestingly, a non Brillouin function type temperature dependence of the ordered magnetic moment has been found. This may arise due to the possible presence of an in-plane geometrical spin frustration in the studied system due to its peculiar crystal structure. Here we would like to point out that for a classical Heisenberg Kagomé lattice antiferromagnet, the ground state was theoretically predicted to be a 120º spin structure, where the vector sum of three spins on the basic triangle is zero [24]. However, the plane of the basic triangle, formed by spins, has the freedom of twist with respect to the planes of the neighbouring triangles. Therefore, the ground state has an infinite and continuous degeneracy, and no long-range ordering can be expected even at zero temperature [25]. In agreement, out results show only a short-range magnetic ordering at lower temperatures. For other classical Kagomé systems, such as, SrCr$_{8-x}$Ga$_{4+x}$O$_{19}$ [26], (D$_3$O)Fe$_3$-(SO$_4$)$_2$(OD)$_6$ jarosite [27, 28] and Li$_2$Mn$_2$O$_4$ [29], broad/diffuse magnetic peaks (corresponding to short-range spin-spin correlations) were also reported over a broad temperature region down to 1.5 K. For these compounds, the diffuse magnetic scattering was reported to be Warren type (saw tooth type) due to a two dimensional magnetic ordering within the Kagomé plane. However, theoretical studies predicted that a thermal or a quantum fluctuation may resolve the degeneracy of the ground state and can result a magnetic long-range ordering in a Kagomé lattice [30, 31]. In addition, the presence of spin vacancies may result a long-range magnetic ordering as experimentally reported for the $(D_3O)Fe_{3-x}Al_x(OD)_6(SO_4)_2$ jarosite compounds [32]. For the present compound, the presence of short-range spin-spin correlations was reported earlier at 10 K [1, 5]. However, the nature of the magnetic correlation, the temperature variation of the correlation length, and the coupled magneto-structural properties were not reported. These properties are essential for a better understanding of the magnetic ground state in the present compound. In the present study, we have carried out a detailed neutron diffraction study as a function of temperature followed by a quantitative data analysis to understand the above points (presented later). Now we discuss the possible magnetic structure for the present compound. Since, the crystal structure of the present compound consists of alternating layers of Kagomé and triangular lattices (stacked along the crystallographic $c$ axis), the spin arrangements in these layers are expected to be different. Now, questions arise if (i) both layers are magnetically ordered or not, and (ii) how strong is the magnetic coupling between layers along the $c$ axis. If both layers are magnetically ordered, the magnetic contributions of individual layers may appear at separate or same $Q$ positions in a diffraction pattern depending on the spin arrangements in the individual layers. For the present compound, the observed broad magnetic peak can be indexed as (1/2 0 2) with respective to the crystallographic hexagonal unit cell. In general, for a Kagomé lattice, two types of magnetic structures have been proposed [24], namely, (i) $q_0$ structure with a uniform chirality (having three sublattices) and (ii) the staggered chiral structure with a larger unit cell of $\sqrt{3} \times \sqrt{3}$ (having nine sublattices). The $Q$ position ($\sim 1.35$ Å$^{-1}$) of the observed magnetic peak in the present study rules out the $q_0$ structure. For the $q_0$ magnetic structure, the strongest peak is expected to appear at $Q = |q_0| = |a^*(1, 0)| = 1.158$ Å$^{-1}$, where, $a^* = (2\pi/a)(2/\sqrt{3})$. This indicates that there is neither an AFM next-nearest neighbor coupling nor a ferromagnetic (FM) coupling to third nearest neighbors that would stabilize the $q_0$ structure [12]. Instead, the observed peak position coincides with $Q = |2 q_{\sqrt{3}}|$

= |2$a$\*(1/3, 1/3)| = 1.34 Å$^{-1}$, a reciprocal lattice vector of the $\sqrt{3}\times\sqrt{3}$ structure indicating that the magnetic structure of the Kagomé plane is a staggered chiral ($\sqrt{3}\times\sqrt{3}$) type. In fact, both quantum and classical theories for the Kagome lattice predict that the larger $\sqrt{3}\times\sqrt{3}$ structure is more favorable than the $q_0$ structure [24, 30]. For a triangular lattice, the magnetic structure may be so called 120º structure [5]. For this spin structure, magnetic reflections are expected at the $Q$ points of ($h$/3 $h$/3 $l$) reflections ($h$/3 = non-integer and $l$ = even) with respect to the hexagonal unit cell [5]. In this case, a super lattice reflection (1/3 1/3 2) is expected at $Q$ position ~ 1.40 Å$^{-1}$. In the present study, the (1/3 1/3 2) reflection is absent. Now, the question arises if the magnetic layers are coupled along the $c$ axis or not. If the magnetic layers are not coupled, only a 2D magnetic ordering within $ab$ planes can be expected. On the other hand, the presence of coupling between layers would lead to a 3D long-range magnetic ordering.

The dimensionality of magnetic ordering for the present system is discussed below. It may be noted that, a 2D magnetic ordering within the Kagomé plane was reported for the Ca substituted compound $Y_{0.5}Ca_{0.5}BaCo_4O_7$ [12]. It was concluded that the triangular layer remains magnetically disordered down to 1.5 K. When the effective magnetic exchange coupling along the $c$ axis is absent or very weak compared to that in the $ab$ plane, an asymmetric peak profile (saw tooth type), with a sharp rise in the intensity at the 2D Bragg scattering angle $2\theta_B$ and a slow fall at higher scattering angles, is expected in the powder diffraction patterns. The scattered intensity of a 2D Bragg reflection ($hk$), indexed with only two Miller indices, can be expressed by the Warren function [33-35] as,

$$I_{hk}\left(2\theta\right) = C\left[\frac{\xi_{2D}}{\left(\lambda\sqrt{\pi}\right)}\right]^{\frac{1}{2}} j_{hk}\left|F_{hk}\right|^2 \frac{\left(1+\cos^2 2\theta\right)}{2\left(\sin^{\frac{3}{2}}\theta\right)} F\left(a\right)$$

where, $C$ is a scale factor, $\xi_{2D}$ is the 2D spin-spin correlation length within the 2D layer, $\lambda$ is the wavelength of the incident neutrons, $j_{hk}$ is the multiplicity of the 2D reflection ($hk$) with 2D magnetic structure factor $F_{hk}$, and $2\theta$ is the scattering angle. The function $F(a)$ is given by

$$F(a) = \int_0^\infty \exp\left[-\left(x^2 - a\right)^2\right] dx$$ where, $a = \left(2\xi_{2D}\sqrt{\pi}/\lambda\right)\left(\sin\theta - \sin\theta_{2DB}\right)$ and $\theta_{2DB}$ is the Bragg angle for the 2D

($hk$) reflection. It may be noted that the observed peak in the present study can be indexed as (2 0) reflection with respect to the $\sqrt{3} \times \sqrt{3}$ spin structure of the Kagomé lattice. Here, for $\sqrt{3} \times \sqrt{3}$ spin structure, the in-plane lattice parameter of the Kagomé lattice is $a = b \sim 10.8615$ Å. On the other hand, a Lorentzian-type broad peak profile would be expected for a 3D short-range magnetic ordering [34]. The observed magnetic peak profile at 22 K, after subtraction of nuclear background, and the fitted profiles with both individual Warren and Lorentzian functions are shown in Fig. 6. Almost equal quality of fits (the agreement factor $R^2 = 0.57$ and 0.63 for the Lorentzian function and the Warren function, respectively) have been obtained for both functions. The Warren function fitting yields a 2D correlation length $\xi_{2D} = 120(10)$ Å. The Lorentzian function fitting gives a 3D correlation length $\xi_{3D} = 14(2)$ Å. The variation of correlation lengths with temperature is shown in Fig. 6 (b). Here, a decrease in the correlation length with increasing temperature has been observed for both models. The present data quality is inadequate to confirm the magnetic ordering dimensionality. A high resolution neutron diffraction study is required to confirm the actual dimensionality of the magnetic ordering for the present compound. Nevertheless, the present study establishes a short-range magnetic correlation.

Now we discuss possible correlation between magnetic ordering and crystal structure as derived from the temperature dependent neutron diffraction study. Figure 7 shows the temperature dependence of different structural parameters, such as, (a) lattice constants and unit cell volume, (b) Co2-O bond lengths for the Kagomé site ($6c$ site), (c) Co1-O bond lengths for the triangular site ($2a$ site), (d) the direct distances between Co ions within a given $ab$ planes as well as perpendicular direction, and (e) in-plane and out-of-plane Co-O-Co bong angles. Significant changes in all structural parameters have been found across the magnetic ordering temperature ($\sim$ 110 K) suggesting a correlation between magnetic ordering and crystal structure. With decreasing temperature, a continuous decrease of the $c$ value, with a change in slope around 110 K, has been observed down to 22 K. Whereas, an unusual temperature dependence for the lattice parameter $a$ has been found. Here, with decreasing temperature, $a$ value decreases first down to ~ 110 K, then it takes an upturn and shows a peak like nature with a maximum ~ 80 K. This unusual nature is also evident in the unit cell volume vs. temperature curve [Fig. 7(a)]. Within a given Kagomé layer, among the three in-plane Co-O bond lengths (two Co2-O1 and one Co2-O2), a decrease in the bond lengths Co2-O2, whereas, an increase in the other two equivalent bond lengths Co2-O1 below 110 K have been observed [Fig. 7(b)]. The bond length Co2-O3 along the perpendicular direction also decreases with lowering of temperature below 110 K. On the other hand, for the triangular site ($2a$), an increase in both in-plane and out-of-plane bond lengths have been observed [Fig. 7(b)]. The side distances (direct distances between Co ions) of two equilateral triangles in the Kagomé layers show an opposite temperature dependence [Fig. 7(d)]. On the other hand, no change in the value of Co1-Co1 distance (within the triangular plane) is evident over the studied temperature range (22 - 300 K) [Fig. 7(d)]. With decreasing temperature, within a given Kagomé lattice, an increase in the Co2–O1–Co2 bond angle has been observed, whereas, the other bond angle Co2–O2–Co2 remains almost constant. A decrease in the Co1–O3–Co2 bond angle (between Co ions from the adjacent layers along the $c$ axis) has also been observed with a lowering of temperature. The changes in bond lengths and bond angle without any change in direct Co-Co distances along the $c$ axis establish a movement of O3 ion within the given plane. All above observations suggest that the magnetic ordering (110 K) is linked to the onset of the changes in all structural parameters, indicating the presence of a coupling between crystal structure and magnetic ordering in the studied compound.

## 4. Summary and conclusion

In summary, we have prepared single phase polycrystalline sample of $YBaCo_4FeO_7$. A detailed crystal and magnetic structural study has been carried out over the temperature range of 22–300 K by neutron diffraction. Rietveld refinement suggests that the system crystallizes in the hexagonal symmetry with the space group $P6_3mc$. The AFM ordering below ~ 110 K has been concluded from the appearance of addition magnetic peak in the neutron diffraction study. At magnetic ordering temperature ~ 110 K, a weak slope change in both temperature dependent $M_{ZFC}$ and $M_{FC}$ curves is found. The linear behaviors of $M(H)$ down to 5 K confirming the presence of an AFM type spin-spin correlations. It has been found that the magnetic correlation remains short-range down to 22 K (lowest measured temperature) along with a non Brillouin function like temperature dependence. The spin structure of the Kagomé layers is found to be staggered chiral ($\sqrt{3} \times \sqrt{3}$) type. The absence of a long-range magnetic ordering in the present compound suggests an uncoupling of Kagomé layers along the $c$ axis via triangular layer. The crystal structural parameters show an observable change around ~ 110 K (magnetic ordering temperature) indicating a coupling between crystal structure and magnetic ordering.

## Acknowledgment

AKB acknowledges the help rendered by A. B. Shinde in performing neutron diffraction experiments.

**Table 1.** The Rietveld refined lattice constants ($a$ and $c$), unit cell volume, fractional atomic coordinates, isotropic thermal parameters, site occupancies, and agreement factors for the sample $YBaCo_4FeO_7$ at 300 and 22 K.

| | $T$ = 300 K | $T$ = 22 K |
|---|---|---|
| $a$ (Å) | 6.2719(5) | 6.2707(6) |
| $c$ (Å) | 10.2060(9) | 10.1701(9) |
| $V$ (Å$^3$) | 347.69(5) | 346.33(6) |
| Space Group | $P6_3mc$ | $P6_3mc$ |
| **Y** | | |
| $2b$ (2/3, 1/3, $z$) | | |
| $z/c$ | 0.8471(3) | 0.8472(4) |
| $B_{iso}$ (Å$^2$) | 0.7(2) | 0.5(2) |
| $Occ.$ | 1.0 | 1.0 |
| **Ba** | | |
| $2b$ (2/3, 1/3, $z$) | | |
| $z/c$ | 0.4680(3) | 0.4729(4) |
| $B_{iso}$ (Å$^2$) | 1.6(1) | 1.2(2) |
| $Occ.$ | 1.0 | 1.0 |
| **Co1** | | |
| $2a$ (0, 0, $z$) | | |
| $z/c$ | 0.4202(2) | 0.4186(5) |
| $B_{iso}$ (Å$^2$) | 1.2(3) | 0.8(2) |
| $Occ.$ | 1.0 | 1.0 |
| **Co2** | | |
| $6c$ ($x$, -$x$, $z$) | | |
| $x/a$ | 0.1764(9) | 0.1725(5) |
| $y/b$ | 0.8235(9) | 0.8297(5) |
| $z/c$ | 0.6668(1) | 0.6618(3) |
| $B_{iso}$ (Å$^2$) | 0. 6(1) | 0.3(2) |
| $Occ.$ | 1.0 | 1.0 |
| **O1** | | |
| $6c$ ($x$, -$x$, $z$) | | |
| $x/a$ | 0.5042(8) | 0.5033(2) |
| $y/b$ | 0.4957(8) | 0.4967(2) |
| $z/c$ | 0.7275(9) | 0.7237(5) |
| $B_{iso}$ (Å$^2$) | 2.8(8) | 2.2(7) |
| $Occ.$ | 1.0 | 1.0 |
| **O2** | | |
| 2*a* (0, 0, *z*) | | |
| $z/c$ | 0.2318(1) | 0.2221(3) |
| $B_{iso}$ (Å$^2$) | 1.0(1) | 0.7(2) |
| *Occ.* | 1.0 | 1.0 |
| | | |
| **O3** | | |
| 6*c* (*x*, -*x*, *z*) | | |
| $x/a$ | 0.1641(1) | 0.1673(5) |
| $y/b$ | 0.8359(2) | 0.8326(4) |
| $z/c$ | 0.4680(9) | 0.4683(3) |
| $B_{iso}$ (Å$^2$) | 3.1(9) | 2.6(5) |
| *Occ.* | 1.0 | 1.0 |
| | | |
| $\chi^2$ | 2.34% | 2.85% |
| $R_p$ | 3.15% | 3.61% |
| $R_{wp}$ | 4.03% | 4.57% |
| $R_{exp}$ | 2.68% | 2.71% |

**Table 2.** The values of bond lengths at 300 K for the compound YBaCo$_4$O$_7$.

| Atom | **O1** | **O2** | **O3** |
| --- | --- | --- | --- |
| **Y** | 3×2.148(8) | – | 3×2.215(7) |
| **Ba** | 3×3.184(14)<br>3×3.077(14) | – | 6×3.136(7) |
| **Co1 (2*a* site)** | – | 1.92(2) | 3×1.849(9) |
| **Co2 (6*c* site)** | 2×1.890(8) | 2.028(10) | 2.034(15) |

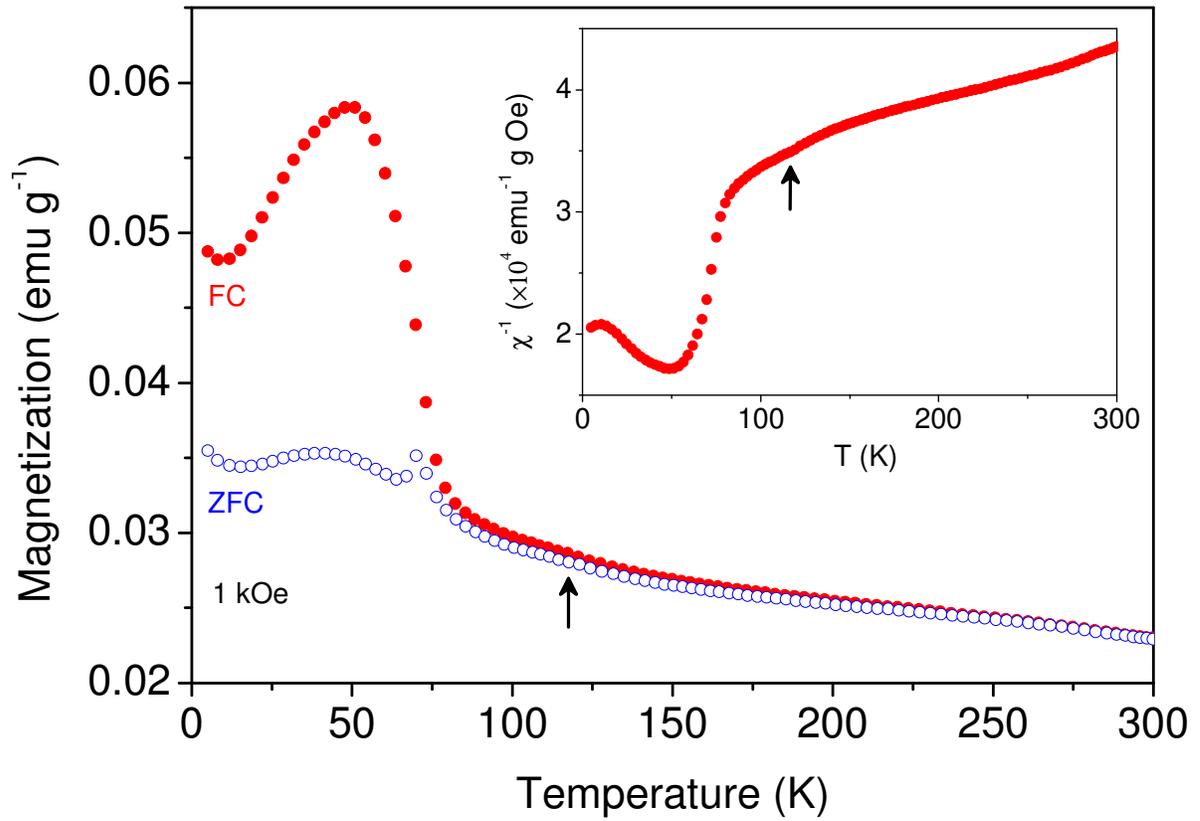

**Fig. 1.** (Color online) Temperature dependence of $M_{ZFC}$ and $M_{FC}$, measured under 1 kOe field, for

YBaCo$_4$O$_7$ compound. Inset shows the $\chi^{-1}(T)$ curve (calculated from the $M_{FC}$ curve).

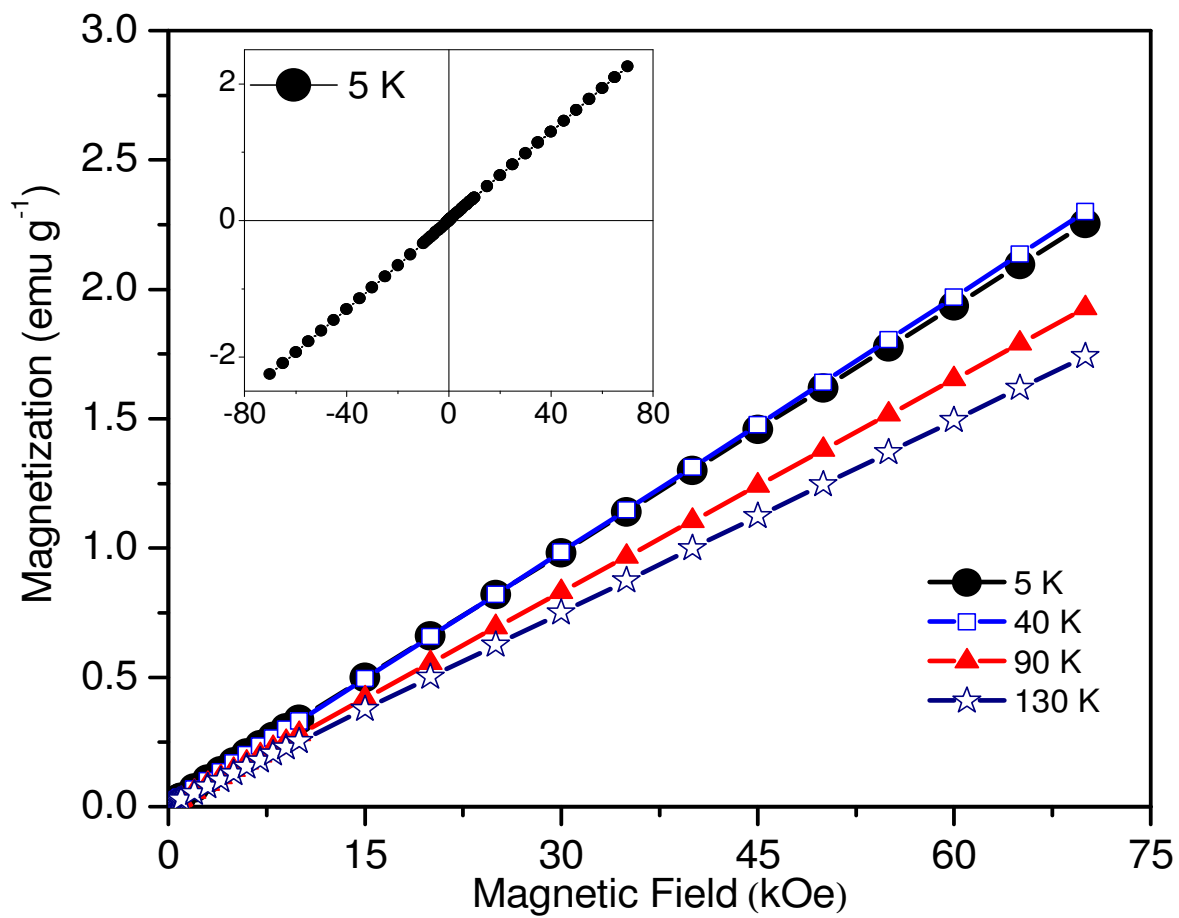

**Fig. 2.** (Color online) The magnetization curves as a function of magnetic field *H* measured at 5, 40, 90, and 130 K. Inset shows the *M*(*H*) curve at 5 K over all four quadrants.

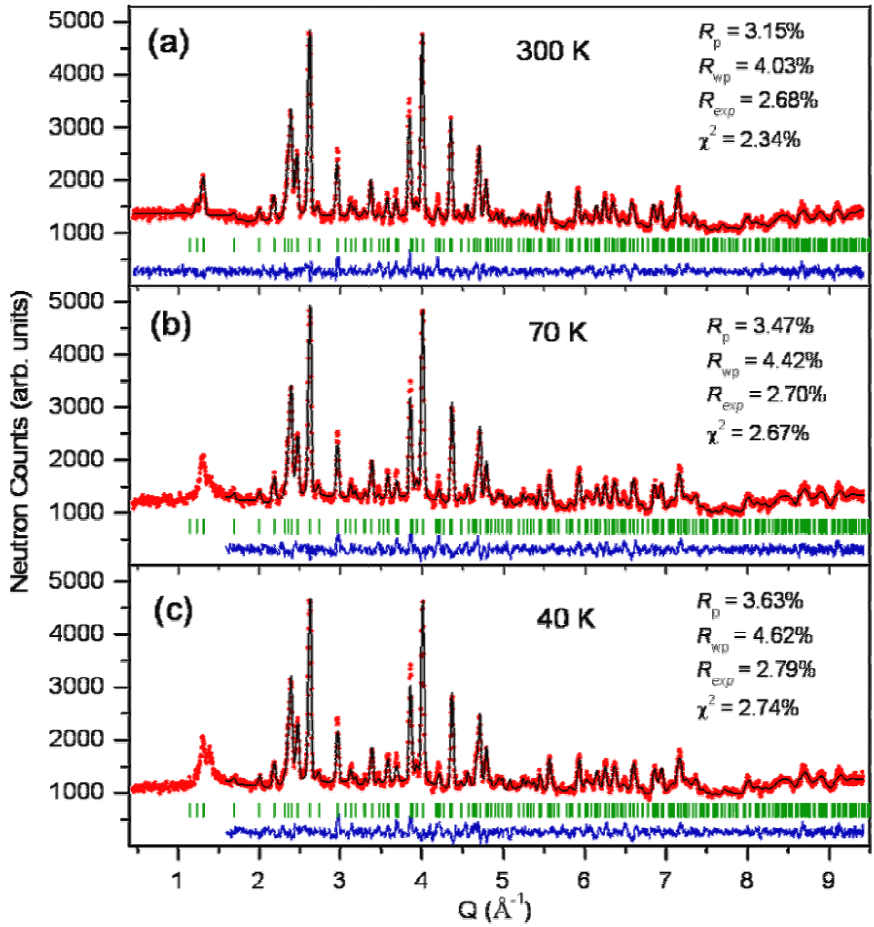

**Fig. 3.** (Color online) Neutron diffraction patterns at (a) 300, (b) 70, and (c) 40 K for the compound $YBaCo_4O_7$. The experimental data are shown by solid circles and the calculated curves, obtained by the Rietveld fitting with a hexagonal ($P6_3mc$) crystal structure, are shown by solid lines. For 70 and 40 K patterns, the lower $Q$ region (0.4-1.6 Å$^{-1}$) has been omitted in the analysis to ensure that the magnetic reflections are not included. Solid line at the bottom of each panel shows the difference between observed and calculated patterns. Vertical lines show the position of the Bragg peaks. The agreement factors such as, profile factor ($R_p$), weighted profile factor ($R_{wp}$), expected weighted profile factor ($R_{exp}$), and $\chi^2$ for all temperatures are also given.

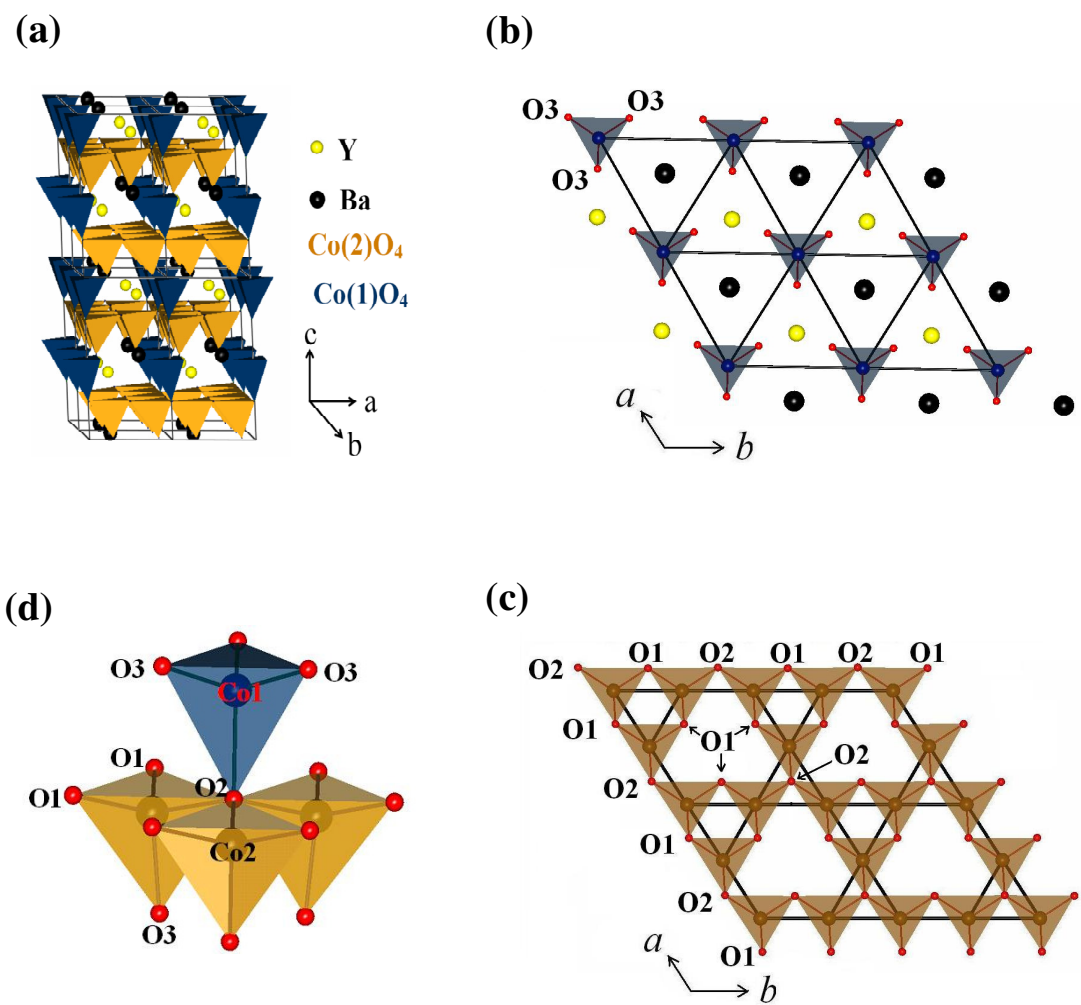

**Fig. 4.** (Color online) (a) The layered type crystal structure of the YBaCo$_4$O$_7$ compound. (b) and (c) The geometrical arrangements of the CoO$_4$ tetrahedra within a given *ab* plane for the 2*a* site (triangular lattice) and 6c site (Kagomé lattice), respectively. (d) Local crystal structure showing the stacking of the triangular and Kagomé layers along the *c* axis.

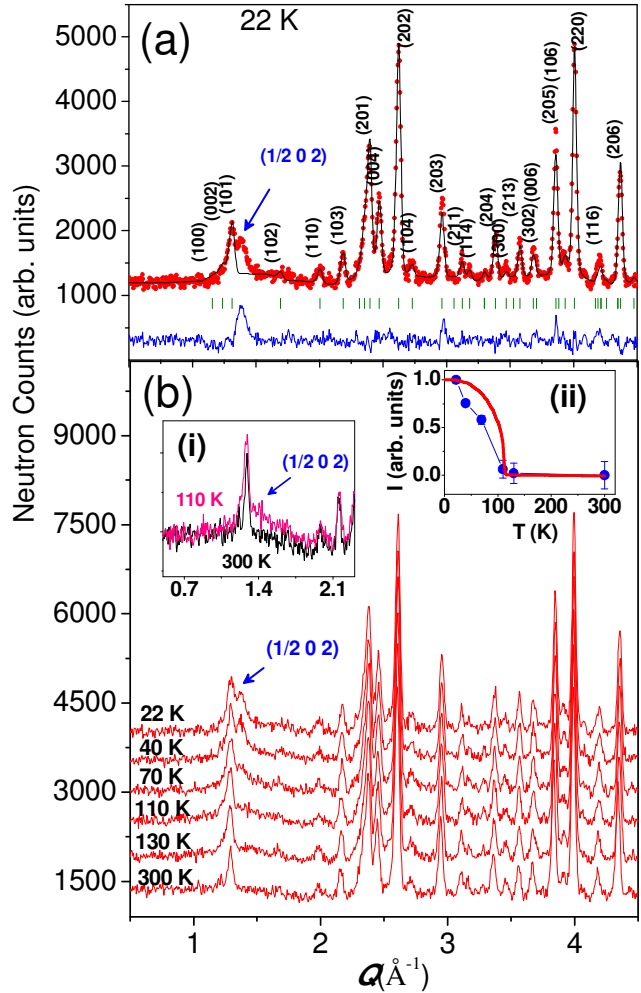

**Fig. 5.** (Color online) (a) Experimentally (circles) observed neutron diffraction patterns at 22 K and a calculated (solid line) curve by considering only nuclear phase (crystal structure). Solid line at the bottom shows the difference between observed and calculated patterns. Vertical lines show the position of the nuclear (fundamental) Bragg peaks. The (*hkl*) values for the observed peaks are also listed. (b) The neutron diffraction patterns measured at 22, 40, 70, 110, 130, and 300 K. The left inset (i) shows a selected range of the neutron diffraction patterns measured at 110 and 300 K. The appearance of the (1/2 0 2) peak at 110 K is evident. The right inset (ii) shows the temperature variation of the integrated intensity of the magnetic peak at $Q$ = 1.35 Å$^{-1}$. The red thick curve shows the calculated intensities according to the Brillouin function. Error bars on the data for the temperatures 70, 40, and 22 K are smaller than the symbol size.

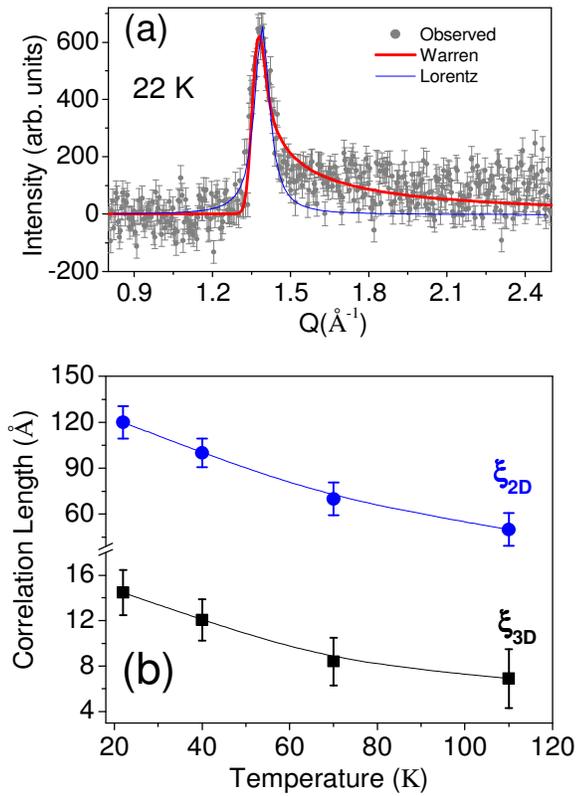

**Fig. 6.** (Color online) (a) The pure magnetic diffraction pattern (after substruction of the nuclear background at 130 K) at 22 K. The thick and thin curves are the calculated profiles according to the Warren function (for a 2D magnetic ordering) and a Lorentzian functions (for a 3D magnetic ordering), respectively. (b) The variation of the 2D and 3D spin-spin correlation lengths as a function of temperature.

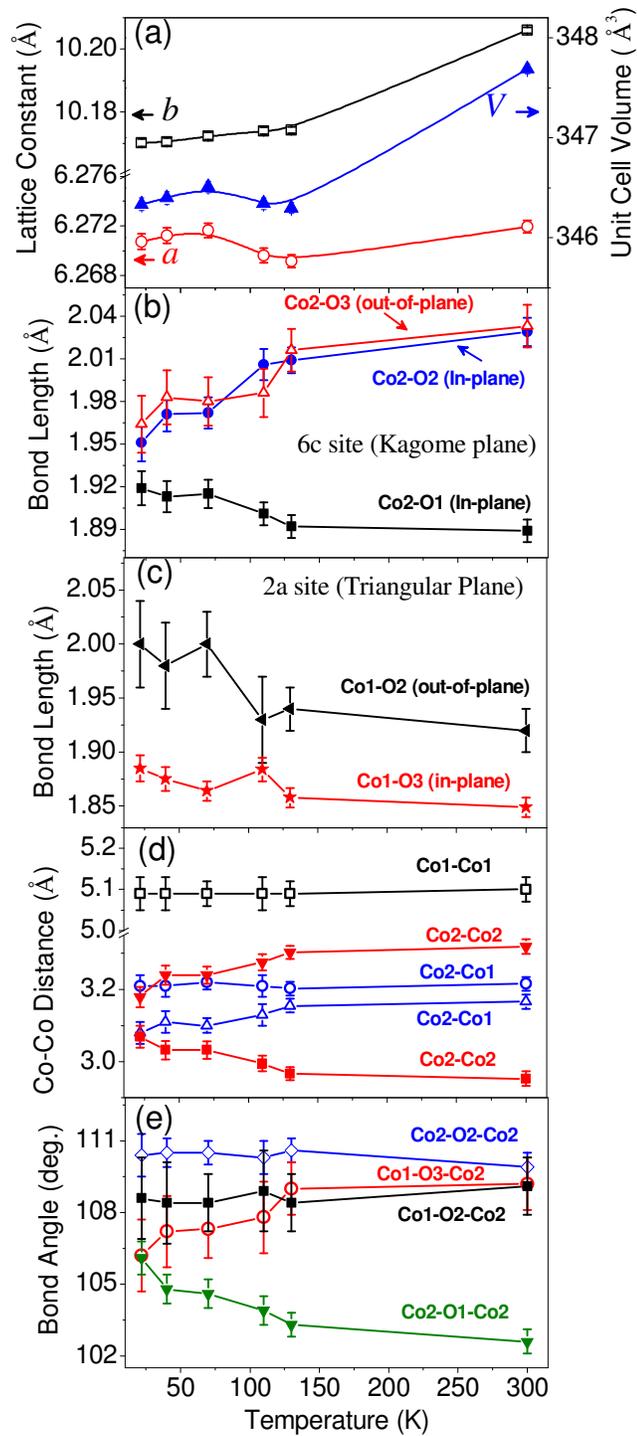

**Fig. 7.** (Color online) The temperature dependence of the (a) lattice constants and unit cell volume, (b) Co2-O bond lengths at Kagomé layer (6c site), (c) Co1-O bond lengths at triangular layer (2*a* site), (d) direct Co-Co distances at both sites, and (e) Co-O-Co bong angles.